\documentclass[nofootinbib,a4paper,letterpaper,superscriptaddress,
twocolumn,showpacs,times,eqsecnum]{revtex4}
\usepackage{amsmath}
\usepackage{amsfonts}
\usepackage{booktabs}
\usepackage{multirow}
\usepackage{siunitx}
\usepackage{adjustbox}
\pdfoutput=1
\usepackage{graphicx}
\usepackage{color}
\usepackage{braket}
\usepackage{dcolumn}
\usepackage{bm,url}
\usepackage[linktocpage]{hyperref}
\usepackage{subfigure}
\usepackage{amsfonts}
\usepackage[usenames,dvipsnames,svgnames]{xcolor}  %% colored text
\usepackage{hyperref}
\definecolor{oxfordblue}{rgb}{0.0, 0.13, 0.28}
\definecolor{burgundy}{rgb}{0.5, 0.0, 0.13}
\definecolor{darkolivegreen}{rgb}{0.33, 0.42, 0.18}
\definecolor{darkblue}{rgb}{0,0,0.5}
\definecolor{richcarmine}{rgb}{0.84, 0.0, 0.25}
\definecolor{darkblue}{rgb}{0,0,0.5}
\definecolor{bluer}{rgb}{0.00,0.50,0.75}{}
\hypersetup{colorlinks=true, citecolor=red, linkcolor=blue,
 urlcolor = magenta, filecolor=magenta}

\usepackage[toc]{appendix}

\usepackage{bm}
\usepackage{amsfonts}
\usepackage{latexsym}
\usepackage{graphicx}

\usepackage{textcomp}
\usepackage{float}
\usepackage{booktabs}
\usepackage{dcolumn}
\usepackage{ragged2e}

\begin{document}
\title{Observational constraints on scalar-nonmetricity cosmology: the role of
the connection }
\author{Ghulam Murtaza
%\orcidlink{0009-0002-6086-7346}
}
\email{ghulammurtaza@1utar.my}
\affiliation{Department of Mathematical and Actuarial Sciences, Universiti Tunku
Abdul Rahman, Jalan Sungai Long,
43000 Cheras, Malaysia}
\author{Avik De
%\orcidlink{0000-0001-6475-3085}
}
\email{avikde@um.edu.my}
\affiliation{Institute of Mathematical Sciences, Faculty of Science, Universiti
Malaya, 50603 Kuala Lumpur, Malaysia}

\author{Emmanuel N. Saridakis\,
%\orcidlink{0000-0003-1500-0874 }
}
\email{msaridak@noa.gr}
\affiliation{Institute for Astronomy, Astrophysics, Space Applications and
Remote Sensing, National Observatory of Athens, 15236 Penteli, Greece}
\affiliation{
Departamento de Matem\'{a}ticas,
Universidad Cat\'{o}lica del Norte, Avda. Angamos 0610, Casilla 1280
Antofagasta, Chile}
\affiliation{CAS Key Laboratory for Researches in Galaxies and
Cosmology, Department of Astronomy, University of Science and Technology of
China, Hefei, Anhui 230026, P.R. China}

\begin{abstract}
Scalar-nonmetricity gravity provides a natural scalar-tensor extension of
symmetric teleparallel gravity, with the gravitational interaction encoded
in nonmetricity. A distinctive feature is that, even for a spatially flat
Friedmann-Robertson-Walker geometry, three inequivalent
symmetric-teleparallel connection branches are allowed. Whether this
geometrical freedom has observable consequences is therefore an important
question. In this work we perform a systematic observational comparison of
all three connection branches, considering a steep scalar-field potential
and a power-law nonminimal coupling. We constrain the models using
combinations of Pantheon Plus and DES Year 5 Type Ia supernovae, DESI DR2
baryon acoustic oscillations, and compressed cosmic microwave background
information. We find that all three branches can reproduce the transition
from matter domination to late-time acceleration, but their observational
behavior is   connection dependent. Connection I exhibits the
strongest sensitivity to the supernova compilation, Connection II yields
remarkably stable reconstructed histories, while Connection III shows an
intermediate behavior. Most notably, for DES Y5+CMB+DESI DR2, Connection I
gives $\Delta\chi^2_{\min}=-13.99$ and $\Delta{\rm AIC}=-5.99$ relative to
$\Lambda$CDM, and is therefore favored by the Akaike information criterion.
\end{abstract}

\maketitle

\section{Introduction}

The observed late-time acceleration of the Universe constitutes one of
the main challenges of modern cosmology and has stimulated an extensive
investigation of   theories beyond the standard $\Lambda$CDM
paradigm \cite{CANTATA:2021asi,Capozziello:2011et}. A natural possibility is
to modify the gravitational sector itself, and a large variety of extensions
of General Relativity (GR) have been developed along this direction. Within
the standard curvature formulation, one may generalize the Einstein-Hilbert
action by introducing nonlinear or higher-order curvature invariants,
leading, among others, to $f(R)$ gravity
\cite{Starobinsky:1980te,Capozziello:2002rd,DeFelice:2010aj},
$f(G)$ gravity \cite{Nojiri:2005jg,DeFelice:2008wz}, cubic gravity
\cite{Asimakis:2022mbe}, and Lovelock gravity
\cite{Lovelock:1971yv,Deruelle:1989fj}. An alternative and particularly
important class is provided by scalar-tensor constructions, including
Horndeski gravity
\cite{Horndeski:1974wa,Santos:2023eqp,Santos:2024cvx} and generalized
Galileon theories \cite{DeFelice:2010nf}. On the other hand, since gravity
admits
an equivalent torsional formulation, one can construct modified theories
starting from teleparallel gravity instead. This leads to $f(T)$ gravity
\cite{Cai:2015emx,Linder:2010py,Chen:2010va}, $f(T,T_G)$ gravity
\cite{Kofinas:2014owa,Kofinas:2014daa}, $f(T,B)$ gravity
\cite{Bahamonde:2015zma,Bahamonde:2016grb}, scalar-torsion theories
\cite{Geng:2011aj}, as well as the Teleparallel Equivalent of Horndeski
gravity
\cite{Bahamonde:2019shr,Bahamonde:2020cfv,Capozziello:2023foy,
Santos:2024zoh,DosSantos:2022exb}.

A third, geometrically equivalent, description of gravity is provided by
symmetric teleparallel gravity, originally discussed by
Nester~\cite{Nester:1998mp}. In this formulation both curvature and torsion
vanish, while the gravitational interaction is entirely encoded in
nonmetricity. This alternative geometrical representation has opened a new
direction for constructing modified gravity theories, with $f(Q)$ gravity
being its simplest and most extensively studied extension
\cite{BeltranJimenez:2017tkd,BeltranJimenez:2019tme}. In this case the
nonmetricity scalar $Q$ appearing linearly in the symmetric teleparallel
equivalent of GR is promoted to a general function $f(Q)$. The resulting
theory exhibits a rich phenomenology and has been investigated in a variety
of cosmological and astrophysical applications, including background
cosmology, observational constraints, and static spherically symmetric
configurations
\cite{Anagnostopoulos:2021ydo,Lazkoz:2019sjl,Lu:2019hra,
Mandal:2020buf,Frusciante:2021sio,Gadbail:2022jco,Khyllep:2021pcu,
Barros:2020bgg,Shabani:2023xfn,De:2023xua,Dimakis:2021gby,
Anagnostopoulos:2022gej,Guzman:2023oyl,Blixt:2023kyr,
Guzman:2024cwa,Wang:2024eai,Boiza:2025xpn,Alwan:2025nka,
Basilakos:2025olm,Li:2025msm}. For a   review of nonmetricity-based gravity and
its extensions, see \cite{Heisenberg:2023lru}.

Following the same logic that leads from General Relativity to scalar-tensor
gravity, or
from teleparallel gravity to scalar-torsion theories, one may extend
symmetric teleparallel gravity by introducing a scalar field nonminimally
coupled to the nonmetricity scalar
\cite{Berthias:1993aa,Jarv:2018bgs,Bahamonde:2022esv,Runkla:2018xrv, Jarv:2023sbp,
Paliathanasis:2023gfq, Dimakis:2023cam,
Murtaza:2026n,Flathmann:2021itp,Gadbail:2024lek,Dimakis:2024fan,Mavoa:2026xzt,
Panotopoulos:2025ajd,Gohain:2023vhb,
Chen:2025mhu,
GhulamMurtaza:2026phx,Paliathanasis:2024sle,
Paliathanasis:2024tit,Subramaniam:2025, Murtaza:2025gra}. Scalar-nonmetricity
gravity therefore
provides the natural nonmetricity counterpart of the familiar
scalar-curvature and scalar-torsion constructions. Apart from enlarging the
space of viable cosmological behaviors, such a framework has an additional
feature which is absent in the usual metric scalar-tensor theories. Namely,
the affine connection is an independent geometrical quantity, and
cosmological symmetry does not select a unique symmetric-teleparallel
connection. For a spatially flat Friedmann-Robertson-Walker geometry there
exist three inequivalent connection branches, which lead to different
expressions for the nonmetricity scalar and, in general, to different
background cosmological equations
\cite{Hohmann:2020zre,Hohmann:2021ast,DAmbrosio:2021pnd,
Paliathanasis:2023nkb,Paliathanasis:2015aos,Dimakis:2016mip,
Heisenberg:2022mbo, Shi:2023kvu}. Hence, the choice of connection is
not
merely a technical issue, but may leave observable signatures in the cosmic
expansion history.

An additional ingredient of considerable interest is the form of the
scalar-field potential. Although the exponential potential is frequently
employed because of its mathematical simplicity, more general potentials
can lead to a substantially richer evolution
\cite{WaliHossain:2014usl}. In particular, potentials that
become steeper than the standard exponential form naturally introduce a
varying effective slope and may generate nontrivial transitions during the
cosmological evolution \cite{Shahalam:2017rit,Das:2019ixt}. At first sight,
steep potentials might appear unfavorable for late-time
acceleration, since a rapidly
varying potential tends to enhance the scalar-field evolution. Nevertheless,
this conclusion can be significantly modified in the presence of a
nonminimal gravitational coupling
\cite{Faraoni:2000wk,Skugoreva:2013ooa}. In scalar-nonmetricity gravity the
interplay between the steepness of the potential, the scalar-nonmetricity
coupling, and the affine connection may therefore lead to viable late-time
cosmologies. Determining whether such scenarios are actually supported by
observations, and whether the three allowed connection branches are
observationally distinguishable, is thus an important question.

In the present work we address this question by confronting
scalar-nonmetricity cosmology with current observational data. We consider
a steep scalar potential and a power-law nonminimal coupling and analyze
separately all three symmetric-teleparallel connection branches.
The dynamical properties of scalar-nonmetricity cosmologies with steep
potentials have recently been investigated in \cite{Murtaza:2025zoo}, while a
data-driven reconstruction of the dynamical connection function in
$f(Q)$ cosmology was performed in \cite{Yang:2024tkw}. However, a systematic
observational comparison of the three symmetric-teleparallel connection
branches in scalar-nonmetricity cosmology has not yet been performed.
In the present work we fill this gap by confronting all three branches
with current observational data.

 In particular, we derive the background evolution equations in
a form suitable for numerical integration and perform a Markov Chain Monte
Carlo analysis using combinations of the Pantheon Plus and DES Year 5
Type Ia supernova samples, DESI DR2 baryon acoustic oscillation measurements,
and compressed cosmic microwave background information. We obtain
constraints on the cosmological and model parameters, reconstruct the
evolution of the matter and effective dark-energy density parameters and
of the effective equation of state, and investigate the correlations
between the parameters. Additionally, we compare the statistical
performance of the three connection branches with that of $\Lambda$CDM.

The manuscript is organized as follows. In Sec. \ref{framework} we present the
scalar-nonmetricity framework and derive the cosmological equations for the
three symmetric-teleparallel connection branches. In Sec. \ref{background} we
rewrite
the corresponding background equations in terms of dimensionless variables
and redshift in the form required for the observational analysis. In
Sec. \ref{observations} we introduce the observational datasets and methodology,
present
the resulting parameter constraints and reconstructed cosmological
evolution, and perform the posterior and model-comparison analysis.
Finally, Sec. \ref{Conclusion} is devoted to the conclusions.

%%%%%%%%%%%%%%%%%%%%%%%%%%%%%%%%%%%%%%%%%%%%%%%%%%%%%%%%%%%%%%%%%%%%%%%%%%%%%%%%
%
\section{Scalar-tensor symmetric teleparallel cosmology}
\label{framework}

In this section we present the scalar-tensor extension of symmetric
teleparallel gravity considered in the present work and derive the
corresponding cosmological equations. We first introduce the general
scalar-nonmetricity framework and obtain the gravitational, scalar-field,
and connection field equations. We then specialize to a spatially flat
Friedmann-Robertson-Walker geometry and examine the three independent
symmetric-teleparallel connection branches allowed by cosmological
symmetry.

\subsection{Scalar-nonmetricity gravity}
\label{subsec:theory}

In symmetric teleparallel gravity, the gravitational interaction is encoded in
nonmetricity, while both curvature and torsion of the affine connection vanish.
More specifically, alongside the metric $g_{\mu\nu}$ one considers an
independent torsion-free and curvature-free affine connection
$\Gamma^\alpha{}_{\mu\nu}$. The latter can be decomposed as
\begin{equation}
\Gamma^\lambda{}_{\mu\nu}
=
\mathring{\Gamma}^\lambda{}_{\mu\nu}
+
L^\lambda{}_{\mu\nu},
\end{equation}
where $\mathring{\Gamma}^\lambda{}_{\mu\nu}$ is the Levi-Civita connection,
\begin{equation}
\mathring{\Gamma}^\alpha{}_{\mu\nu}
=
\frac{1}{2}g^{\alpha\beta}
\left(
\partial_\mu g_{\beta\nu}
+
\partial_\nu g_{\beta\mu}
-
\partial_\beta g_{\mu\nu}
\right),
\end{equation}
and $L^\lambda{}_{\mu\nu}$ is the disformation tensor. The nonmetricity tensor
is defined as
\begin{equation}
Q_{\lambda\mu\nu}
\equiv
\nabla_\lambda g_{\mu\nu}
=
\partial_\lambda g_{\mu\nu}
-
\Gamma^\beta{}_{\lambda\mu}g_{\beta\nu}
-
\Gamma^\beta{}_{\lambda\nu}g_{\beta\mu},
\end{equation}
and the disformation reads
\begin{equation}
L^\lambda{}_{\mu\nu}
=
\frac{1}{2}
\left(
Q^\lambda{}_{\mu\nu}
-
Q_\mu{}^\lambda{}_\nu
-
Q_\nu{}^\lambda{}_\mu
\right).
\end{equation}
Introducing the two nonmetricity vectors
$Q_\mu\equiv Q_\mu{}^\nu{}_\nu$ and
$\widetilde Q_\mu\equiv Q_{\nu\mu}{}^\nu$, the nonmetricity conjugate is
\begin{equation}
P^\lambda{}_{\mu\nu}
=
\frac{1}{4}
\left[
-2L^\lambda{}_{\mu\nu}
+
\left(Q^\lambda-\widetilde Q^\lambda\right)g_{\mu\nu}
-
\delta^\lambda{}_{(\mu}Q_{\nu)}
\right],
\end{equation}
while, with the conventions adopted in the present work, the nonmetricity scalar
is
\begin{equation}
Q=Q_{\alpha\beta\gamma}P^{\alpha\beta\gamma}.
\end{equation}

A scalar-tensor extension of symmetric teleparallel gravity is obtained by
allowing a scalar field to couple nonminimally to the nonmetricity scalar
\cite{Jarv:2018bgs}. We consider the action
\begin{eqnarray}
&&
\!\!\!\!\!\!\!\!\!\!\!
S=
\frac{1}{2\kappa}
\int d^4x\,\sqrt{-g}
\Big[
f(\phi)Q
-
h(\phi)\nabla^\mu\phi\nabla_\mu\phi
-
U(\phi)\Big]\nonumber\\
&& \
+ S_m,
\label{action}
\end{eqnarray}
with $\kappa$ the gravitational constant, and
where $f(\phi)$ determines the nonminimal coupling, $h(\phi)$ specifies the
kinetic sector, and $U(\phi)$ is the scalar-field potential, while $S_m$
accounts for the matter action. This construction
constitutes the symmetric-teleparallel counterpart of curvature-based
scalar-tensor gravity. In particular, the choice $f(\phi)=\phi$ and
$h(\phi)=\omega/\phi$ leads to the nonmetricity analogue of Brans-Dicke
gravity. Moreover, in the vanishing-kinetic-term limit the theory reproduces
$F(Q)$ gravity through the usual auxiliary-field representation, with
$\phi=F_Q$ and $U(\phi)=F_Q Q-F(Q)$.

Variation of (\ref{action}) with respect to the metric gives
\begin{align}
\kappa T_{\mu\nu}
&=
f\,\mathring G_{\mu\nu}
+
2f_{,\phi}P^\lambda{}_{\mu\nu}\nabla_\lambda\phi
-
h\nabla_\mu\phi\nabla_\nu\phi
\nonumber\\
&\ \ \ +
\frac{1}{2}h g_{\mu\nu}
\nabla^\alpha\phi\nabla_\alpha\phi
+
\frac{1}{2}U g_{\mu\nu},
\label{metricFE}
\end{align}
where $\mathring G_{\mu\nu}$ is the Einstein tensor constructed from the
Levi-Civita connection. Variation with respect to the scalar field yields
\begin{equation}
f_{,\phi}Q
+
h_{,\phi}\nabla^\alpha\phi\nabla_\alpha\phi
+
2h\,\mathring{\Box}\phi
-
U_{,\phi}
=0.
\label{scalarFE}
\end{equation}
Finally, the independent affine connection provides an additional field
equation. For matter minimally coupled to the metric, for which the
hypermomentum vanishes, it may be written in the form
\begin{equation}
\nabla_\mu\nabla_\nu
\left(
\sqrt{-g}\,
f(\phi)P^{\mu\nu}{}_{\lambda}
\right)=0.
\label{connectionFE}
\end{equation}
Hence, although the symmetric teleparallel connection is curvature-free and
torsion-free, its allowed cosmological form is not arbitrary: it must satisfy
(\ref{connectionFE}) together with the metric and scalar-field equations.

In the following we take $h(\phi)=h_0={\rm const}$, which, for a
nonvanishing kinetic sector of fixed sign, can be achieved through an
appropriate scalar-field redefinition. Furthermore, motivated by the
possibility of obtaining a richer scalar-field evolution than that associated
with the standard exponential potential, we consider
\begin{eqnarray}
&&U(\phi)
=
U_0
\exp\left[
\alpha
\left(
\frac{\phi}{M_{\rm Pl}}
\right)^n
\right],\\
&&
f(\phi)=\beta\phi^m ,
\label{model_functions}
\end{eqnarray}
where $U_0$, $\alpha$, $\beta$, $m$, and $n$ are constants. The standard
exponential potential is recovered for $n=1$, while $n>1$ corresponds to the
steep-potential case considered in the present work. In what follows we set the
Planck mass as
$M_{\rm Pl}^2\equiv 1/\kappa=1$.

 \subsection{FRW cosmology and connection branches}
\label{subsec:cosmology}

We proceed to the cosmological application of the theory. We consider a
spatially flat, homogeneous and isotropic Friedmann-Robertson-Walker (FRW)
geometry,
\begin{equation}
ds^2
=
-dt^2
+
a^2(t)
\left(
dx^2+dy^2+dz^2
\right),
\label{FRWmetric}
\end{equation}
where $a(t)$ is the scale factor and
$H\equiv\dot a/a$ is the Hubble parameter. The scalar field is homogeneous,
$\phi=\phi(t)$, while the matter sector is described by a perfect fluid,
\begin{equation}
T_{\mu\nu}
=
p\,g_{\mu\nu}
+
(\rho+p)u_\mu u_\nu .
\end{equation}

Unlike in the metric formulation of gravity, specifying the FRW metric does
not uniquely determine the affine connection. Cosmological symmetry must be
imposed independently on the flat and torsion-free connection. For a
spatially flat FRW geometry, the most general symmetric-teleparallel
connection compatible with homogeneity and isotropy separates into three
branches
\cite{Hohmann:2020zre,Hohmann:2021ast,DAmbrosio:2021pnd,
Paliathanasis:2023nkb,Paliathanasis:2015aos,Dimakis:2016mip,
Heisenberg:2022mbo,Shi:2023kvu}.
In Cartesian coordinates its nonvanishing time-dependent components can be
parametrized as
\begin{equation}
\Gamma^t{}_{tt}=C_1(t),
\ \
\Gamma^t{}_{ij}=C_2(t)\delta_{ij},
\ \
\Gamma^i{}_{tj}
=
\Gamma^i{}_{jt}
=
C_3(t)\delta^i{}_j .
\label{general_connection}
\end{equation}
Imposing the vanishing-curvature condition then leads to the three families
\begin{align}
{\rm I}: \quad&
C_1=\gamma(t),
\qquad
C_2=C_3=0,
\label{connI}
\\
{\rm II}: \quad&
C_1=
\gamma(t)+\frac{\dot\gamma(t)}{\gamma(t)},
\qquad
C_2=0,
\qquad
C_3=\gamma(t),
\label{connII}
\\
{\rm III}: \quad&
C_1=-\frac{\dot\gamma(t)}{\gamma(t)},
\qquad
C_2=\gamma(t),
\qquad
C_3=0,
\label{connIII}
\end{align}
where $\gamma(t)\neq0$ for Connections II and III.

It is useful to clarify the role of $\gamma(t)$. It parametrizes the residual
freedom of the flat and torsion-free affine connection and should not, by
itself, be regarded as an additional matter degree of freedom. Its evolution
is constrained by the connection equation (\ref{connectionFE}). Moreover,
the three branches do not affect the cosmological evolution in the same way.
For Connection I the function $\gamma(t)$ drops out from the background
equations, and the resulting background cosmology is equivalent to that
obtained in the usual coincident gauge. On the contrary, for Connections II
and III the connection function enters both the nonmetricity scalar and the
background field equations, and therefore contributes nontrivially to the
cosmological evolution.

For the parametrization (\ref{general_connection}), the nonmetricity scalar
takes the form
\begin{eqnarray}
&&Q=
3\Big[
-2H^2
+3HC_3
+\frac{HC_2}{a^2}
-\frac{(C_1+C_3)C_2}{a^2}\nonumber\\
&& \ \ \ \ \ \ \ \ \ \
+(C_1-C_3)C_3
\Big].
\label{Qgeneral}
\end{eqnarray}
Using Eqs.~(\ref{connI})-(\ref{connIII}), one obtains respectively
\begin{align}
Q_{\rm I}
&=-6H^2 ,
\label{QI}
\\
Q_{\rm II}
&=-6H^2+9\gamma H+3\dot\gamma ,
\label{QII}
\\
Q_{\rm III}
&=-6H^2
+\frac{3\gamma H}{a^2}
+\frac{3\dot\gamma}{a^2}.
\label{QIII}
\end{align}
Thus, Connection I reproduces the standard FRW expression of the
coincident-gauge formulation, whereas Connections II and III induce explicit
connection-dependent contributions.

For later convenience, the two independent metric equations can be written
for all three branches in the compact form
\begin{align}
\kappa\rho
={}&
3fH^2
+
\frac{3}{2}\dot f
\left(
C_3-\frac{C_2}{a^2}
\right)
-
\frac{1}{2}h_0\dot\phi^2
-
\frac{1}{2}U ,
\label{friedmann1}
\\
\kappa p
={}&
-f\left(2\dot H+3H^2\right)
+
\frac{1}{2}\dot f
\left(
3C_3+\frac{C_2}{a^2}-4H
\right)\nonumber\\
&
-
\frac{1}{2}h_0\dot\phi^2
+
\frac{1}{2}U .
\label{friedmann2}
\end{align}
Similarly, the homogeneous scalar-field equation becomes
\begin{equation}
f_{,\phi}Q
-
2h_0
\left(
\ddot\phi+3H\dot\phi
\right)
-
U_{,\phi}
=0 .
\label{scalarcosmo}
\end{equation}
For minimally coupled perfect-fluid matter, the connection equation leads
to the standard conservation relation
\begin{equation}
\dot\rho+3H(\rho+p)=0.
\label{continuity}
\end{equation}

Equations~(\ref{friedmann1})-(\ref{continuity}), supplemented by the
corresponding connection equation, form the background cosmological system.
Since the three allowed connections lead to different expressions for $Q$
and, for Connections II and III, to an explicit dependence on $\gamma(t)$,
they define three distinct cosmological branches of the same underlying
scalar-nonmetricity theory. In the next section we formulate the
corresponding background equations in a form suitable for their subsequent
confrontation with observations.

 \section{Background equations for observational analysis}
\label{background}

The dynamical properties of scalar-nonmetricity cosmology with the steep
potential and power-law coupling considered here have been investigated in
detail in Ref.~\cite{Murtaza:2025zoo}, including the finite critical points,
their
stability, the asymptotic structure at infinity, and the nonhyperbolic
solutions through center-manifold techniques. In the present work we do not
repeat this analysis. Instead, our purpose is to formulate the background
cosmological equations in the form required for the subsequent confrontation
with observations.

We restrict ourselves to pressureless matter, $p=0$, and set $h_0=1$.
For all three connection branches we introduce the dimensionless quantities
\begin{eqnarray}
&&x=\frac{\dot{\phi}}{\sqrt{6f}\,H},
\qquad
y=\frac{U}{3fH^2},
\qquad
\tilde{\Omega}_m=\frac{\kappa\rho}{3fH^2},
\nonumber\\
&&
\lambda=\frac{U'\sqrt{f}}{U},
\qquad
\zeta=\frac{f'}{\sqrt{f}}.
\label{common_variables}
\end{eqnarray}
For Connections II and III an additional variable associated with the
affine connection will be introduced below. Notice that
$\tilde{\Omega}_m$ is normalized by the nonminimal coupling $f(\phi)$
and therefore does not coincide in general with the standard matter
density parameter.

For the model functions
\begin{equation}
U(\phi)=U_0 e^{\alpha\phi^n},
\qquad
f(\phi)=\beta\phi^m,
\label{model_functions_background}
\end{equation}
one obtains
\begin{equation}
\lambda
=
n\alpha\sqrt{\beta}\,
\phi^{n-1+\frac{m}{2}},
\qquad
\zeta
=
m\sqrt{\beta}\,
\phi^{\frac{m}{2}-1}.
\end{equation}
Hence, for $m\neq2$, these two quantities are not independent. Introducing
\begin{equation}
D\equiv2n-2+m,
\qquad
p_\zeta\equiv\frac{D}{m-2},
\end{equation}
we obtain
\begin{equation}
\lambda={\cal K}\zeta^{p_\zeta},
\qquad
{\cal K}
=
n\alpha\,
m^{-\frac{D}{m-2}}
\beta^{-\frac{n}{m-2}}.
\label{lambda_zeta_relation}
\end{equation}
Thus, the background evolution can be formulated without treating
$\lambda$ as an independent variable. In the following we restrict
ourselves to the positive-field branch, $\phi>0$, and consider
$n>1$, $m>2$, $\alpha>0$, and $\beta>0$, corresponding to the
steep-potential sector employed in the observational analysis.

For each connection we introduce an effective dark-energy density parameter
according to
\begin{equation}
\tilde{\Omega}_{\rm DE}=1-\tilde{\Omega}_m,
\end{equation}
while the total effective equation-of-state parameter is
\begin{equation}
w_{\rm eff}
=
-1-\frac{2}{3}\frac{\dot H}{H^2}.
\label{weff_general}
\end{equation}
For the numerical integration it is convenient to employ the redshift
$z$, with $1+z=a^{-1}$. Therefore
\begin{equation}
\frac{d}{dN}=-(1+z)\frac{d}{dz},
\qquad
N=\ln a.
\label{N_z_relation}
\end{equation}
Finally, in the following, a subscript $0$ denotes the present-day value of the
corresponding quantity, namely its value at $z=0$.
%%%%%%%%%%%%%%%%%%%%%%%%%%%%%%%%%%%%%%%%%%%%%%%%%%%%%%%%%%%%%%%%%%%%%%%%%%%%%%%

\subsection{Connection I}
\label{subsec:connectionI}

For Connection I the affine function $\gamma(t)$ drops out from the
background equations and the nonmetricity scalar reduces to
$Q=-6H^2$. The first Friedmann equation gives
\begin{equation}
\tilde{\Omega}_m
=
1-x^2-\frac{y}{2},
\label{constraint_1}
\end{equation}
and hence
\begin{equation}
\tilde{\Omega}_{\rm DE}
=
x^2+\frac{y}{2}.
\label{OmegaDE_conI}
\end{equation}
The second Friedmann equation yields
\begin{equation}
\frac{\dot H}{H^2}
=
-\frac{3}{2}
-\sqrt{6}\,x\zeta
-\frac{3}{2}x^2
+\frac{3}{4}y,
\label{Hdot_conI}
\end{equation}
from which we deduce
\begin{equation}
w_{\rm eff}
=
\frac{2\sqrt{2}}{\sqrt{3}}x\zeta
+x^2-\frac{y}{2}.
\label{weff_conI}
\end{equation}
Since matter is pressureless, the corresponding effective dark-energy
equation of state can also be written as
\begin{equation}
w_{\rm DE}
=
\frac{w_{\rm eff}}{\tilde{\Omega}_{\rm DE}}
=
\frac{
\frac{2\sqrt{2}}{\sqrt{3}}x\zeta+x^2-\frac{y}{2}
}{
x^2+\frac{y}{2}
},
\label{wDE_conI}
\end{equation}
whenever $\tilde{\Omega}_{\rm DE}\neq0$.

Using Eq.~(\ref{lambda_zeta_relation}) to eliminate $\lambda$, the
background system to be integrated in redshift takes the compact form
\begin{align}
\frac{dx}{dz}
&=
-\frac{1}{4(1+z)}
\left[
-\left(\sqrt{6}{\cal K}\zeta^{p_\zeta}+3x\right)y\right.\nonumber\\
&\left. \ \ \ \ \ \ \ \ \ \ \ \ \ \ \ \ \ \ \
+2(x^2-1)(3x+\sqrt{6}\zeta)
\right],
\label{xz_conI}
\\
\frac{dy}{dz}
&=
-\frac{y}{2(1+z)}
\left[
6+6x^2-3y
+2\sqrt{6}\,x
\left({\cal K}\zeta^{p_\zeta}+\zeta\right)
\right],
\label{yz_conI}
\\
\frac{d\zeta}{dz}
&=
-\frac{\sqrt{6}}{1+z}
\frac{m-2}{2m}\,
x\zeta^2,
\label{zetaz_conI}
\\
\frac{dH}{dz}
&=
\frac{H}{1+z}
\left[
\frac{3}{2}
+\sqrt{6}\,x\zeta
+\frac{3}{2}x^2
-\frac{3}{4}y
\right].
\label{Hz_conI}
\end{align}
Together with Eqs.~(\ref{constraint_1}) and
(\ref{OmegaDE_conI}), these equations completely determine the
background evolution of Connection I and provide the system used in the
observational analysis of Sec. \ref{observations}.

\subsection{Connection II}
\label{subsec:connectionII}

For Connection II the affine connection contributes explicitly to the
background evolution. In addition to the variables introduced in
Eq.~(\ref{common_variables}), we define
\begin{equation}
\eta=\frac{\gamma}{H}.
\label{eta_conII}
\end{equation}
The first Friedmann equation then gives
\begin{align}
\tilde{\Omega}_m
&=
1+\sqrt{\frac{3}{2}}\,x\eta\zeta
-x^2-\frac{y}{2},
\label{constraint_conII}
\\
\tilde{\Omega}_{\rm DE}
&=
x^2+\frac{y}{2}
-\sqrt{\frac{3}{2}}\,x\eta\zeta,
\label{OmegaDE_conII}
\end{align}
while for pressureless matter, the second Friedmann equation yields
\begin{equation}
\frac{\dot H}{H^2}
=
-\frac{3}{2}
+\frac{3}{2}\sqrt{\frac{3}{2}}\,x\eta\zeta
-\sqrt{6}\,x\zeta
-\frac{3}{2}x^2
+\frac{3}{4}y.
\label{Hdot_conII}
\end{equation}
Hence, the total effective equation-of-state parameter is
\begin{equation}
w_{\rm eff}
=
-\sqrt{\frac{3}{2}}\,x\eta\zeta
+\frac{2\sqrt{2}}{\sqrt{3}}\,x\zeta
+x^2-\frac{y}{2}.
\label{weff_conII}
\end{equation}
Since the matter component is pressureless, the effective dark-energy
equation of state is simply
$w_{\rm DE}=w_{\rm eff}/\tilde{\Omega}_{\rm DE}$ whenever
$\tilde{\Omega}_{\rm DE}\neq0$.

Using the model-specific relation $\lambda={\cal K}\zeta^{p_\zeta}$
from Eq.~(\ref{lambda_zeta_relation}), the background equations can
be written directly in terms of the redshift as
\begin{align}
\frac{dx}{dz}
={}&
-\frac{1}{1+z}
\Bigg[
-\frac{3}{2}x
+\frac{3}{2}x^3
-\frac{3}{4}xy
-\sqrt{\frac{3}{2}}\,
\frac{m-2}{m}\,x^2\zeta\nonumber\\
& \ \ \ \ \ \ \ \ \ \ \  \ \
-\frac{3}{2}\sqrt{\frac{3}{2}}\,
x^2\eta\zeta
\Bigg],
\label{xz_conII}
\\
\frac{dy}{dz}
={}&
-\frac{y}{1+z}
\Bigg[
3
+\sqrt{6}\,{\cal K}\zeta^{p_\zeta}x
+3x^2
-\frac{3}{2}y
+\sqrt{6}\,x\zeta \nonumber\\
& \ \ \ \ \ \ \ \ \ \ \  \ \
-3\sqrt{\frac{3}{2}}\,x\eta\zeta
\Bigg],
\label{yz_conII}
\\
\frac{d\eta}{dz}
={}&
-\frac{1}{1+z}
\Bigg[
2-\frac{3}{2}\eta
+\frac{3}{2}x^2\eta
-\frac{3}{4}y\eta
-\frac{4(m-1)}{m}x^2
\nonumber\\
&
+{\cal K}y\zeta^{p_\zeta-1}
+\sqrt{6}\,x\eta\zeta
-\frac{3}{2}\sqrt{\frac{3}{2}}\,
x\eta^2\zeta
\Bigg],
\label{etaz_conII}
\\
\frac{d\zeta}{dz}
={}&
-\frac{\sqrt{6}}{1+z}
\frac{m-2}{2m}\,
x\zeta^2,
\label{zetaz_conII}
\\
\frac{dH}{dz}
={}&
\frac{H}{1+z}
\left[
\frac{3}{2}
-\frac{3}{2}\sqrt{\frac{3}{2}}\,x\eta\zeta
+\sqrt{6}\,x\zeta
+\frac{3}{2}x^2
-\frac{3}{4}y
\right].
\label{Hz_conII}
\end{align}
Notice that the term $\lambda/\zeta$ appearing in the unreduced
formulation is now replaced by
${\cal K}\zeta^{p_\zeta-1}$. Since $m>2$ and $n>1$ imply
$p_\zeta>1$, this quantity remains regular in the limit
$\zeta\rightarrow0$.

Equations~(\ref{xz_conII})-(\ref{Hz_conII}), together with the constraint
(\ref{constraint_conII}), constitute the system used for the numerical
background evolution and the observational analysis of Connection II.

 \subsection{Connection III}
\label{subsec:connectionIII}

For Connection III we introduce
$\mathring{\gamma}\equiv\gamma/a^2$ and define the dimensionless
connection variable
\begin{equation}
\eta=\frac{\mathring{\gamma}}{H}.
\label{eta_conIII}
\end{equation}
The first Friedmann equation then gives
\begin{align}
\tilde{\Omega}_m
&=
1-\sqrt{\frac{3}{2}}\,x\eta\zeta
-x^2-\frac{y}{2},
\label{constraint_conIII}
\\
\tilde{\Omega}_{\rm DE}
&=
x^2+\frac{y}{2}
+\sqrt{\frac{3}{2}}\,x\eta\zeta,
\label{OmegaDE_conIII}
\end{align}
while for pressureless matter, the second Friedmann equation becomes
\begin{equation}
\frac{\dot H}{H^2}
=
-\frac{3}{2}
-\frac{3}{2}x^2
+\frac{3}{4}y
-\sqrt{6}\,x\zeta
+\frac{1}{2}\sqrt{\frac{3}{2}}\,x\eta\zeta.
\label{Hdot_conIII}
\end{equation}
Consequently, the total effective equation-of-state parameter is
\begin{equation}
w_{\rm eff}
=
x^2-\frac{y}{2}
+\frac{x\zeta(4-\eta)}{\sqrt{6}}.
\label{weff_conIII}
\end{equation}
Since the matter sector is pressureless, the corresponding effective
dark-energy equation of state is
$w_{\rm DE}=w_{\rm eff}/\tilde{\Omega}_{\rm DE}$ whenever
$\tilde{\Omega}_{\rm DE}\neq0$.

For compactness, we further introduce
\begin{equation}
\Xi\equiv4\sqrt{6}\,x+3\eta\zeta.
\label{Xi_conIII}
\end{equation}
Using the model-specific relation
$\lambda={\cal K}\zeta^{p_\zeta}$ from
Eq.~(\ref{lambda_zeta_relation}), the background equations can be written
directly in terms of the redshift as
\begin{align}
\frac{dx}{dz}
={}&
-\frac{3x}{4(1+z)\Xi}
\Bigg\{
8\sqrt{6}\,x^3
-8{\cal K}\zeta^{p_\zeta}y
-2x^2(-8+\eta)\zeta
\nonumber\\
&\hspace{1.5cm} \ \ \ \ \
+\left[(10-3y)\eta-16\right]\zeta\nonumber\\
&
-\sqrt{6}\,x
\left[
4(2+y)
+\eta\left(\eta+2-\frac{4}{m}\right)\zeta^2
\right]
\Bigg\},
\label{xz_conIII}
\\
\frac{dy}{dz}
={}&
-\frac{y}{2(1+z)}
\Big\{
6+6x^2-3y\nonumber\\
& \ \ \ \ \ \ \ \ \ \ \ \ \  \ \ \
+\sqrt{6}\,x
\left[
2{\cal K}\zeta^{p_\zeta}
+(2-\eta)\zeta
\right]
\Big\},
\label{yz_conIII}
\\
\frac{d\eta}{dz}
={}&
-\frac{\eta}{4(1+z)\Xi}
\Bigg[12{\cal K}\zeta^{p_\zeta}y
-3\left[-8+3(2+y)\eta\right]\zeta
\nonumber\\
&\hspace{1.5cm} \ \ \ \ +
24\sqrt{6}\,x^3
-6x^2
\left(
\eta-8-\frac{8}{m}
\right)\zeta\nonumber\\
& \ \ \ \ \ \ \
+\sqrt{6}\,x
\left[
8-12y-3(-4+\eta)\eta\zeta^2
\right]
\Bigg],
\label{etaz_conIII}
\\
\frac{d\zeta}{dz}
={}&
-\frac{\sqrt{6}}{1+z}
\frac{m-2}{2m}\,
x\zeta^2,
\label{zetaz_conIII}
\\
\frac{dH}{dz}
={}&
\frac{H}{1+z}
\left[
\frac{3}{2}
+\frac{3}{2}x^2
-\frac{3}{4}y
+\sqrt{6}\,x\zeta
-\frac{1}{2}\sqrt{\frac{3}{2}}\,x\eta\zeta
\right].
\label{Hz_conIII}
\end{align}

The system is well defined for $\Xi\neq0$, and the numerical integrations
performed below are restricted to this regular domain.  Together with the
constraint (\ref{constraint_conIII}), Eqs.~(\ref{xz_conIII})-(\ref{Hz_conIII})
constitute the background system used for the
observational analysis of Connection III.

\section{Observational constraints and cosmological implications}
\label{observations}

In this section we confront the three scalar-nonmetricity connection
branches with current cosmological observations and investigate the extent
to which the choice of symmetric-teleparallel connection affects the
inferred background evolution. We first describe the observational
datasets, likelihoods, and statistical methodology employed in the analysis.
We then present the resulting parameter constraints and reconstruct the
cosmological evolution for each branch. Finally, we compare the statistical
performance of the three
connections with that of the standard $\Lambda$CDM scenario.

\subsection{Observational datasets and methodology}
\label{subsec:data_method}

We constrain the model parameters through a Markov Chain Monte Carlo
(MCMC) analysis implemented within the COBAYA framework
\cite{Torrado:2020dgo}. The background evolution equations derived in
Sec. \ref{background}   are incorporated through CosmoDS \cite{cosmods},
which provides
an external theory component for COBAYA suitable for the numerical
integration of nonstandard cosmological backgrounds. The resulting
posterior samples are analyzed and visualized using the publicly available
GetDist package.

\subsubsection{Type Ia supernovae}

Type Ia supernovae (SN-Ia) constitute one of the main probes of the
late-time expansion history, since their standardized luminosities provide
measurements of the luminosity distance over a broad redshift interval
\cite{reiss1998supernova}. In the present
analysis we consider separately the Pantheon Plus compilation
\cite{Scolnic:2021amr,Riess_2022,Brout:2022vxf} and the DES Year 5
supernova sample \cite{DES:2024jxu}. The two compilations are based on
different surveys, calibration procedures, and analysis pipelines, and
therefore provide complementary determinations of the distance-redshift
relation.

\subsubsection{DESI DR2 BAO}

We additionally employ the baryon acoustic oscillation (BAO) measurements
from the 2025 DESI Data Release 2 (DR2)
\cite{DESI:2025zgx,DESI:2024mwx}. BAO observations provide a standard
ruler through the sound horizon at the drag epoch, $r_d$, and constrain
both radial and transverse cosmological distances. The relevant quantities
are
\begin{equation}
\frac{D_H(z)}{r_d}
=
\frac{c}{H(z)r_d},
\qquad
\frac{D_M(z)}{r_d}
=
\frac{c}{r_d}
\int_0^z \frac{d\tilde z}{H(\tilde z)},
\label{eq:bao_distances}
\end{equation}
where $D_H$ and $D_M$ denote the Hubble and transverse comoving distances,
respectively. For isotropic BAO measurements we use
\begin{equation}
\frac{D_V(z)}{r_d}
=
\left[
\frac{zD_M^2(z)D_H(z)}{r_d^3}
\right]^{1/3}.
\label{eq:DV}
\end{equation}
Throughout the analysis, $r_d$ is treated as a free parameter rather than
being fixed by a specific early-time cosmological model.

\begin{table}[ht]
\centering
\begin{tabular}{c|ccc}
\hline
Parameter & Connection I & Connection II & Connection III \\
\hline
$H_0$       & $[60,80]$       & $[60,80]$       & $[60,80]$ \\
$x_0$       & $[-1,1]$        & $[-0.3,0.3]$    & $[-0.3,0.3]$ \\
$y_0$       & $[10^{-4},2]$   & $[0.5,2]$       & $[0.5,2]$ \\
$\eta_0$    & --              & $[0.05,0.5]$    & $[0.05,0.5]$ \\
$\zeta_0$   & $[0.01,1]$      & $[0.15,0.35]$   & $[0.15,0.35]$ \\
%$\lambda_0$ & $[-0.1,1]$      & $[10^{-6},0.3]$ & $[-0.1,0.3]$ \\
$m$         & $[2.01,3.5]$    & $[2.01,3.5]$    & $[2.01,3.5]$ \\
$n$         & $[1.01,2]$      & $[1.01,2]$      & $[1.01,2]$ \\
$r_d$       & $[130,180]$     & $[130,180]$     & $[130,180]$ \\
$\alpha$    & $1$             & $1$             & $1$ \\
$\beta$     & $1$             & $1$             & $1$ \\
\hline
\end{tabular}
\caption{\textit{Flat prior ranges adopted in the MCMC analysis for the
three symmetric teleparallel connection branches. The parameters
$\alpha$ and $\beta$ are fixed to unity.}}
\label{tab:priors_all}
\end{table}

\begin{table*}
	\centering
	\tiny
	\begin{tabular}{l c c c c c c c c c c}
		\hline
		\multicolumn{8}{c}{\bf Datasets: Pantheon Plus + CMB + DESI DR2}\\
		\hline
		\hline
		Model & $H_0$ & $\widetilde{\Omega}_{m0}$ & $r_d$ & $x_{0}$ & $y_{0}$ &
$\zeta_{0}$ & $m$ & $n$& $\lambda_0$ & $w_{\rm eff,0}$ \\
		\hline
		{\bf Connection I:}
		& $69.40^{+0.51}_{-0.45}$
		& $0.29^{+0.004}_{-0.005}$
		& $146.91^{+0.303}_{-0.261}$
		& $-0.16^{+0.094}_{-0.050}$
		& $1.34^{+0.067}_{-0.019}$
        &$0.15^{+0.029}_{-0.122}$
     %   &$0.28^{+0.174}_{-0.266}$
		& $2.29^{+0.090}_{-0.091}$
        & $1.84^{+0.076}_{-0.092}$
        & $ 1.47^{+1.46}_{-1.44}\times10^{-9}$
		& $-0.69^{+0.014}_{-0.021}$ \\

		{\boldmath\bf $\Lambda$CDM:}
		&$69.53^{+0.35}_{-0.36}$
		& $0.29 \pm 0.004$
		& $146.92^{+0.24}_{-0.25}$
		&
        &
        &
        &
        &
        &
       & $-0.70 \pm 0.004$\\

		\hline
		\multicolumn{8}{c}{\bf Datasets: DES Y5 + CMB + DESI DR2}\\
		\hline
		\hline

		{\bf Connection I:}
		& $67.34^{+0.50}_{-0.54}$
		& $0.31^{+0.005}_{-0.005}$
		& $147.10^{+0.255}_{-0.252}$
		& $-0.25^{+0.040}_{-0.041}$
		& $1.24^{+0.047}_{-0.043}$
        &$0.10^{+0.041}_{-0.067}$
      %  &$>0.72$
		& $2.30^{+0.099}_{-0.101}$
        & $1.84^{+0.104}_{-0.095}$
        & $ 3.83^{+3.83}_{-3.81}\times10^{-10}$
		& $-0.60^{+0.033}_{-0.022}$\\

		{\boldmath\bf $\Lambda$CDM:}
	&$68.63^{+0.42}_{-0.40}$
		& $0.30 \pm 0.004$
		& $147.16^{+0.27}_{-0.23}$
		&
        &
        &
        &
        &
       &
	    &$-0.69 \pm 0.004$  \\

		\hline
		\multicolumn{8}{c}{\bf Datasets: CMB + DESI DR2}\\
		\hline
		\hline

		{\bf Connection I:}
		& $68.23^{+0.89}_{-0.71}$
		& $0.30^{+0.007}_{-0.008}$
		& $147.14^{+0.220}_{-0.266}$
		& $-0.35^{+0.046}_{-0.051}$
		& $1.13^{+0.079}_{-0.073}$
        &$0.41^{+0.065}_{-0.047}$
      %  &$0.13^{+0.105}_{-0.137}$
		& $2.27^{+0.092}_{-0.106}$
        & $1.87^{+0.098}_{-0.099}$
        & $ 4.48^{+3.13}_{-4.45}\times10^{-8}$
		& $-0.68^{+0.023}_{-0.031}$\\

		{\boldmath\bf $\Lambda$CDM:}
		&$69.01^{+0.43}_{-0.45}$
		& $0.29^{+0.004}_{-0.005}$
		& $147.09 \pm 0.25$
		&
        &
        &
        &
       &
       &
	    & $-0.70^{+0.004}_{-0.005}$  \\

		\hline
		\hline
	\end{tabular}
	\caption{Marginalized MCMC constraints for Connection I and $\Lambda$CDM from
Pantheon Plus + CMB + DESI DR2, DES Y5 + CMB + DESI DR2, and CMB + DESI DR2, shown as
median values with 68\% credible intervals.}
	\label{tab:best_fit_valuesI}
\end{table*}

\begin{table*}
	\centering
	\tiny
	\begin{tabular}{l c c c c c c c c c c c}
		\hline
		\multicolumn{8}{c}{\bf Datasets: Pantheon Plus + CMB + DESI DR2}\\
		\hline
		\hline
		Model & $H_0$ & $\widetilde{\Omega}_{m0}$ & $r_d$ & $x_{0}$ & $y_{0}$
&$\eta_{0}$ & $\zeta_{0}$ & $m$ & $n$ & $\lambda_0$ & $w_{\rm eff,0}$\\
		\hline
		{\bf Connection II:}
		& $68.67^{+0.26}_{-0.27}$
		& $0.30^{+0.004}_{-0.003}$
		& $147.24^{+0.238}_{-0.228}$
		& $-0.00010^{+0.00069}_{-0.00067}$
		& $1.39^{+0.007}_{-0.008}$
        & $0.20^{+0.038}_{-0.028}$
        &$0.26^{+0.043}_{-0.028}$
     %   &$<0.068$
		& $2.30^{+0.026}_{-0.030}$
        & $1.40^{+0.038}_{-0.034}$
        & $ 5.94^{+3.84}_{-5.90}\times10^{-9}$
		& $-0.69^{+0.004}_{-0.003}$\\

		{\boldmath\bf $\Lambda$CDM:}
		&$69.53^{+0.35}_{-0.36}$
		& $0.29 \pm 0.004$
		& $146.92^{+0.24}_{-0.25}$
		&
        &
        &
        &
        &
        &
        &
       & $-0.70 \pm 0.004$\\

		\hline
		\multicolumn{8}{c}{\bf Datasets: DES Y5 + CMB + DESI DR2}\\
		\hline
		\hline

		{\bf Connection II:}
		& $68.09^{+0.29}_{-0.30}$
		& $0.30^{+0.004}_{-0.004}$
		& $147.27^{+0.243}_{-0.232}$
		& $-0.00008^{+0.00084}_{-0.00080}$
		& $1.38^{+0.008}_{-0.009}$
        & $0.17^{+0.030}_{-0.016}$
        &$0.23^{+0.009}_{-0.024}$
      %  &$<0.015$
		& $2.18^{+0.015}_{-0.005}$
        & $1.37^{+0.023}_{-0.022}$
        & $ 2.44^{+1.73}_{-2.22}\times10^{-15}$
		& $-0.69^{+0.004}_{-0.004}$\\

		{\boldmath\bf $\Lambda$CDM:}
		&$68.63^{+0.42}_{-0.40}$
		& $0.30 \pm 0.004$
		& $147.16^{+0.27}_{-0.23}$
		&
        &
        &
        &
        &
        &
       &
	    &$-0.69 \pm 0.004$  \\

		\hline
		\multicolumn{8}{c}{\bf Datasets: CMB + DESI DR2}\\
		\hline
		\hline

		{\bf Connection II:}
		& $68.15^{+0.30}_{-0.30}$
		& $0.30^{+0.003}_{-0.004}$
		& $147.25^{+0.256}_{-0.216}$
		& $0.00007^{+0.00069}_{-0.00066}$
		& $1.38^{+0.008}_{-0.007}$
        & $0.17^{+0.042}_{-0.052}$
        &$0.26^{+0.011}_{-0.014}$
      %  &$<0.016$
		& $2.28^{+0.012}_{-0.013}$
        & $1.42^{+0.017}_{-0.020}$
        & $ 1.51^{+1.57}_{-1.22}\times10^{-10}$
		& $-0.69^{+0.004}_{-0.004}$\\

		{\boldmath\bf $\Lambda$CDM:}
	&$69.01^{+0.43}_{-0.45}$
		& $0.29^{+0.004}_{-0.005}$
		& $147.09 \pm 0.25$
		&
        &
        &
        &
        &
       &
       &
	    & $-0.70^{+0.004}_{-0.005}$  \\

		\hline
		\hline
	\end{tabular}
	\caption{Marginalized MCMC constraints for Connection II and $\Lambda$CDM from
Pantheon Plus + CMB + DESI DR2, DES Y5 + CMB + DESI DR2, and CMB + DESI DR2, shown as
median values with 68\% credible intervals.}
	\label{tab:best_fit_valuesII}
\end{table*}

\begin{table*}
	\centering
	\tiny
	\begin{tabular}{l c c c c c c c c c c c}
		\hline
		\multicolumn{8}{c}{\bf Datasets: Pantheon Plus + CMB + DESI DR2}\\
		\hline
		\hline
		Model & $H_0$ & $\widetilde{\Omega}_{m0}$ & $r_d$ & $x_{0}$ & $y_{0}$
&$\eta_{0}$ & $\zeta_{0}$ & $m$ & $n$ & $\lambda_0$ & $w_{\rm eff,0}$\\
		\hline
		{\bf Connection III:}
		& $68.73^{+0.14}_{-0.14}$
		& $0.29^{+0.017}_{-0.016}$
		& $147.20^{+0.214}_{-0.177}$
		& $-0.04624^{+0.04128}_{-0.02415}$
		& $1.41^{+0.033}_{-0.031}$
        & $0.19^{+0.018}_{-0.019}$
        &$0.25^{+0.039}_{-0.023}$
      %  &$0.013^{+0.011}_{-0.011}$
		& $2.31^{+0.011}_{-0.021}$
        & $1.41^{+0.022}_{-0.011}$
        & $ 6.16^{+5.08}_{-4.62}\times10^{-10}$
		& $-0.72^{+0.022}_{-0.017}$\\

		{\boldmath\bf $\Lambda$CDM:}
			&$69.53^{+0.35}_{-0.36}$
		& $0.29 \pm 0.004$
		& $146.92^{+0.24}_{-0.25}$
		&
        &
        &
        &
        &
        &
        &
       & $-0.70 \pm 0.004$\\

		\hline
		\multicolumn{8}{c}{\bf Datasets: DES Y5 + CMB + DESI DR2}\\
		\hline
		\hline

		{\bf Connection III:}
		& $68.63^{+0.17}_{-0.15}$
		& $0.27^{+0.025}_{-0.014}$
		& $147.28^{+0.213}_{-0.250}$
		& $-0.06734^{+0.06618}_{-0.04147}$
		& $1.42^{+0.033}_{-0.036}$
        & $0.17^{+0.053}_{-0.057}$
        &$0.26^{+0.060}_{-0.033}$
     %   &$0.012^{+0.055}_{-0.045}$
		& $2.34^{+0.033}_{-0.024}$
        & $1.41^{+0.040}_{-0.041}$
        & $ 4.74^{+4.13}_{-4.54}\times10^{-9}$
		& $-0.72^{+0.019}_{-0.027}$\\

		{\boldmath\bf $\Lambda$CDM:}
		&$68.63^{+0.42}_{-0.40}$
		& $0.30 \pm 0.004$
		& $147.16^{+0.27}_{-0.23}$
		&
        &
        &
        &
        &
        &
       &
	    &$-0.69 \pm 0.004$  \\

		\hline
		\multicolumn{8}{c}{\bf Datasets: CMB + DESI DR2}\\
		\hline
		\hline

		{\bf Connection III:}
		& $68.56^{+0.15}_{-0.14}$
		& $0.27^{+0.045}_{-0.030}$
		& $147.25^{+0.198}_{-0.207}$
		& $0.01023^{+0.14472}_{-0.14292}$
		& $1.41^{+0.050}_{-0.051}$
        & $0.25^{+0.032}_{-0.061}$
        &$0.24^{+0.0632}_{-0.0473}$
     %   &$-0.036^{+0.039}_{-0.038}$
		& $2.29^{+0.028}_{-0.025}$
        & $1.40^{+0.021}_{-0.022}$
        & $ 1.94^{+1.82}_{-1.88}\times10^{-10}$
		& $-0.68^{+0.014}_{-0.047}$\\

		{\boldmath\bf $\Lambda$CDM:}
		&$69.01^{+0.43}_{-0.45}$
		& $0.29^{+0.004}_{-0.005}$
		& $147.09 \pm 0.25$
		&
        &
        &
        &
        &
       &
       &
	    & $-0.70^{+0.004}_{-0.005}$  \\

		\hline
		\hline
	\end{tabular}
	\caption{Marginalized MCMC constraints for Connection III and $\Lambda$CDM from
Pantheon Plus + CMB + DESI DR2, DES Y5 + CMB + DESI DR2, and CMB + DESI DR2, shown as
median values with 68\% credible intervals.}
	\label{tab:best_fit_valuesIII}
\end{table*}

\subsubsection{Compressed CMB likelihood}

In order to complement the low-redshift probes with information from the
early Universe, we employ the compressed cosmic microwave background
(CMB) likelihood described in Appendix A of
Ref.~\cite{DESI:2025zgx}. The CMB information relevant for the present
background analysis is encoded in a Gaussian likelihood involving the
acoustic angular scale $\theta_\ast$ and the physical baryon and matter
densities,
$\omega_b\equiv\Omega_b h^2$ and
$\omega_{bc}\equiv(\Omega_b+\Omega_c)h^2$.
These quantities are tightly constrained by CMB observations and depend
only weakly on the details of the late-time cosmological evolution once
secondary contributions are appropriately marginalized over
\cite{Lemos:2023xhs}.

The compressed likelihood therefore provides a high-redshift anchor for
the expansion history while avoiding the computational cost and additional
model dependence associated with a full CMB power-spectrum analysis. In
particular, it constrains the matter sector and breaks degeneracies that
remain when BAO and supernova data are considered alone. As shown in
Ref.~\cite{DESI:2025zgx}, for late-time dark-energy analyses the resulting
constraints closely reproduce those obtained with the corresponding full
CMB likelihood.

\subsubsection{Dataset combinations and priors}

Having introduced the three observational probes, we combine their
likelihoods assuming statistical independence. Thus, for the combinations
including supernovae, the total chi-square is
\begin{equation}
\chi^2_{\rm tot}
=
\chi^2_{\rm SN}
+
\chi^2_{\rm BAO}
+
\chi^2_{\rm CMB},
\label{chi2_total}
\end{equation}
while for the CMB+DESI DR2 combination the supernova contribution is
omitted.

We consider the following three combinations of observational datasets:
\begin{enumerate}
    \item Pantheon Plus + CMB + DESI DR2,
    \item DES Y5 + CMB + DESI DR2,
    \item CMB + DESI DR2.
\end{enumerate}
The first two combinations allow us to assess separately the impact of the
Pantheon Plus and DES Y5 supernova compilations, while the third provides a
supernova-independent constraint based on BAO and CMB information.

For each connection branch we sample the present-day values of the
corresponding background variables together with $H_0$, $r_d$, $m$, and
$n$. The parameters $\alpha$ and $\beta$ are fixed to unity throughout
the analysis. Flat priors are adopted within the ranges summarized in
Table~\ref{tab:priors_all}.

\subsection{Observational constraints and cosmological evolution}
\label{subsec:constraints_evolution}

We now present the observational constraints on the three connection
branches and examine the corresponding reconstructed cosmological
evolution. The marginalized constraints obtained from the three dataset
combinations are summarized in Tables~\ref{tab:best_fit_valuesI},
\ref{tab:best_fit_valuesII}, and \ref{tab:best_fit_valuesIII} for
Connections I, II, and III, respectively. In addition to the directly sampled
parameters, we report the derived
present-day quantities $\lambda_0$, $\widetilde{\Omega}_{m0}$, and
$w_{\rm eff,0}$. In particular, $\lambda_0$ is calculated through
Eq.~(\ref{lambda_zeta_relation}) and is therefore not treated as an
independent MCMC parameter.    In all cases the quoted
constraints correspond to the marginalized posterior distributions, with
the uncertainties referring to the $68\%$ credible intervals.

\begin{figure*}[t]
	\centering
	\includegraphics[width=\textwidth]
	{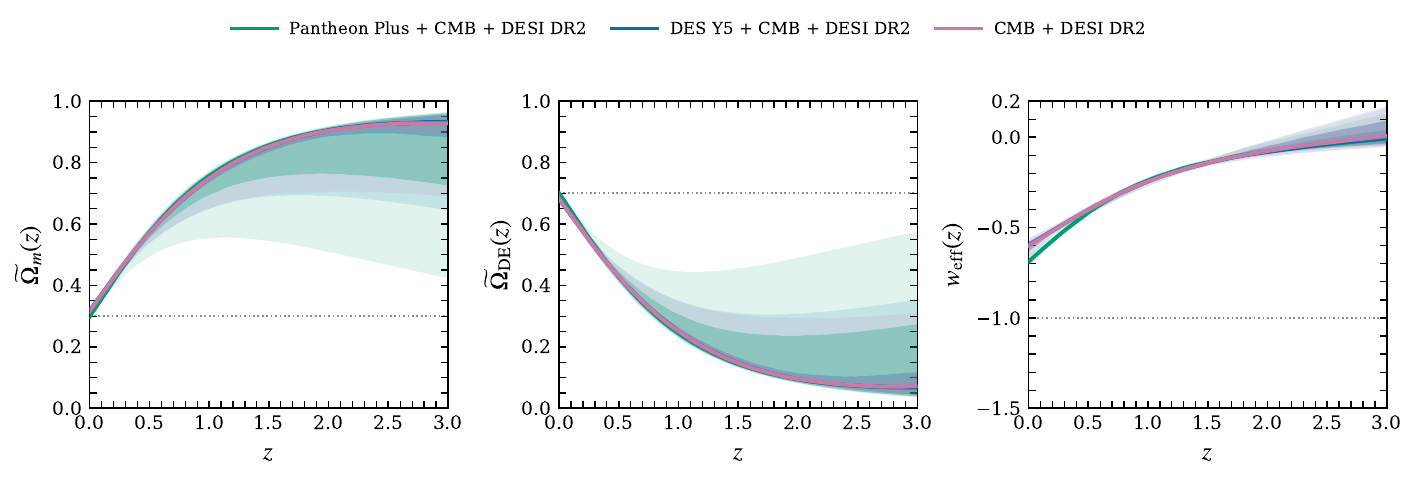}
	\caption{\textit{Reconstructed cosmological evolution for Connection I.
	From left to right, the panels show the normalized matter density parameter
	$\widetilde{\Omega}_{m}(z)$, the effective dark-energy density parameter
	$\widetilde{\Omega}_{\rm DE}(z)$, and the total effective equation-of-state
	parameter $w_{\rm eff}(z)$. The green, blue, and magenta curves correspond,
	respectively, to Pantheon Plus+CMB+DESI DR2, DES Y5+CMB+DESI DR2, and
	CMB+DESI DR2. The solid curves denote the posterior mean evolution,
	while the darker and lighter shaded regions represent the $68\%$ and
	$95\%$ credible intervals, respectively.}}
	\label{fig:evolutionI}
\end{figure*}

\begin{figure*}[t]
	\centering
	\includegraphics[width=\textwidth]
	{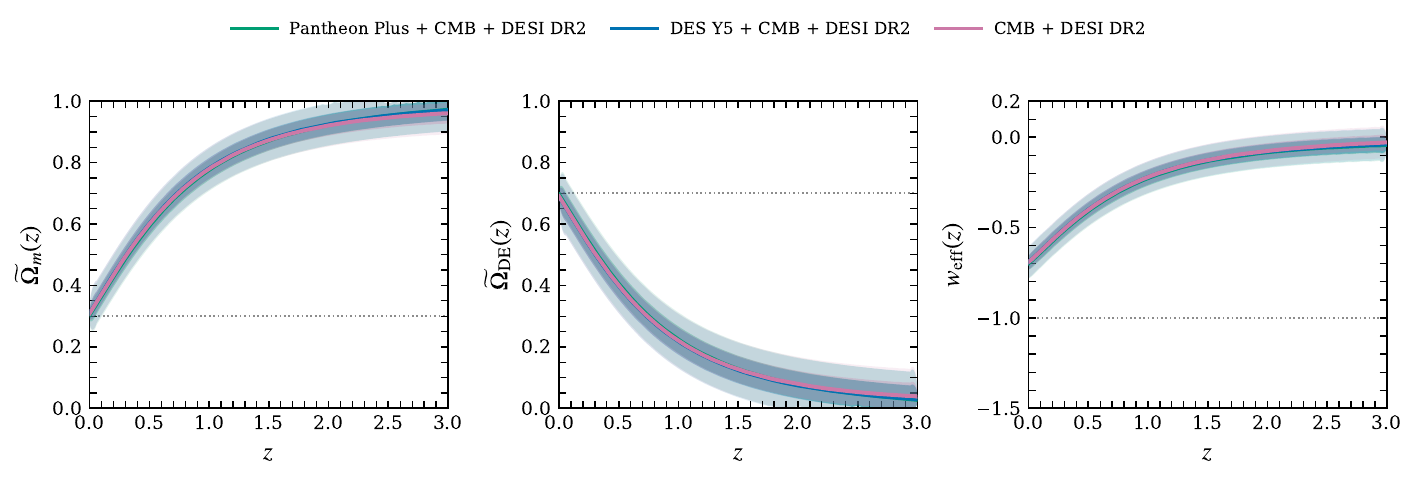}
	\caption{\textit{Reconstructed cosmological evolution for Connection II.
	From left to right, the panels show the normalized matter density parameter
	$\widetilde{\Omega}_{m}(z)$, the effective dark-energy density parameter
	$\widetilde{\Omega}_{\rm DE}(z)$, and the total effective equation-of-state
	parameter $w_{\rm eff}(z)$. The green, blue, and magenta curves correspond,
	respectively, to Pantheon Plus+CMB+DESI DR2, DES Y5+CMB+DESI DR2, and
	CMB+DESI DR2. The solid curves denote the posterior mean evolution, while
	the darker and lighter shaded regions represent the $68\%$ and $95\%$
	credible intervals, respectively.}}
	\label{fig:evolutionII}
\end{figure*}

\begin{figure*}[t]
	\centering
	\includegraphics[width=\textwidth]
	{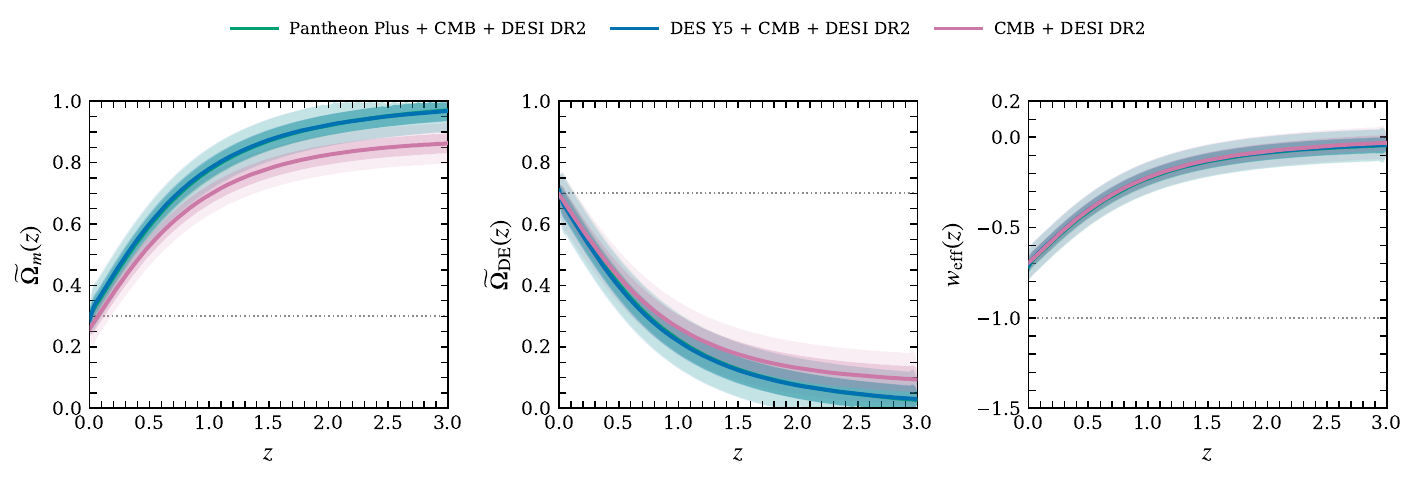}
	\caption{\textit{Reconstructed cosmological evolution for Connection III.
	From left to right, the panels show the normalized matter density parameter
	$\widetilde{\Omega}_{m}(z)$, the effective dark-energy density parameter
	$\widetilde{\Omega}_{\rm DE}(z)$, and the total effective equation-of-state
	parameter $w_{\rm eff}(z)$. The green, blue, and magenta curves correspond,
	respectively, to Pantheon Plus+CMB+DESI DR2, DES Y5+CMB+DESI DR2, and
	CMB+DESI DR2. The solid curves denote the posterior mean evolution, while
	the darker and lighter shaded regions represent the $68\%$ and $95\%$
	credible intervals, respectively.}}
	\label{fig:evolutionIII}
\end{figure*}

For Connection I, the Pantheon Plus+CMB+DESI DR2 combination yields
$H_0=69.40^{+0.51}_{-0.45}\ {\rm km\,s^{-1}\,Mpc^{-1}}$ and
$\widetilde{\Omega}_{m0}=0.29^{+0.004}_{-0.005}$, while the corresponding
sound-horizon scale is
$r_d=146.91^{+0.303}_{-0.261}\ {\rm Mpc}$. The present effective equation
of state is
$w_{\rm eff,0}=-0.69^{+0.014}_{-0.021}$. Replacing Pantheon Plus with DES
Y5 leads to a lower value,
$H_0=67.34^{+0.50}_{-0.54}\ {\rm km\,s^{-1}\,Mpc^{-1}}$, together with
$\widetilde{\Omega}_{m0}=0.31\pm0.005$ and
$w_{\rm eff,0}=-0.60^{+0.033}_{-0.022}$. On the other hand, the
CMB+DESI DR2 combination gives
$H_0=68.23^{+0.89}_{-0.71}\ {\rm km\,s^{-1}\,Mpc^{-1}}$ and
$w_{\rm eff,0}=-0.68^{+0.023}_{-0.031}$. Hence, Connection I displays a
noticeable sensitivity to the adopted supernova compilation, particularly
in the inferred value of $H_0$ and in the present effective equation of
state. This sensitivity is considerably weaker for the other two
connection branches, as we discuss below.

The corresponding reconstructed histories are shown in
Fig.~\ref{fig:evolutionI}. All three dataset combinations recover a
cosmological evolution in which the normalized matter contribution
increases towards the past and the effective dark-energy component becomes
dominant at late times. Correspondingly, $w_{\rm eff}(z)$ evolves from
values close to zero at higher redshift to negative values at the present
epoch, yielding the transition to accelerated expansion. Although the
overall behavior is common to the three reconstructions, their late-time
evolution is not identical. In particular, the DES Y5 combination favors a
less negative present value of $w_{\rm eff}$ and allows a visibly different
late-time evolution, while the Pantheon Plus and CMB+DESI DR2
reconstructions remain closer to each other. Thus, for Connection I the
choice of the supernova dataset has a non-negligible impact on the inferred
recent expansion history.

\begin{figure*}[t]
	\centering
	\includegraphics[width=\textwidth]{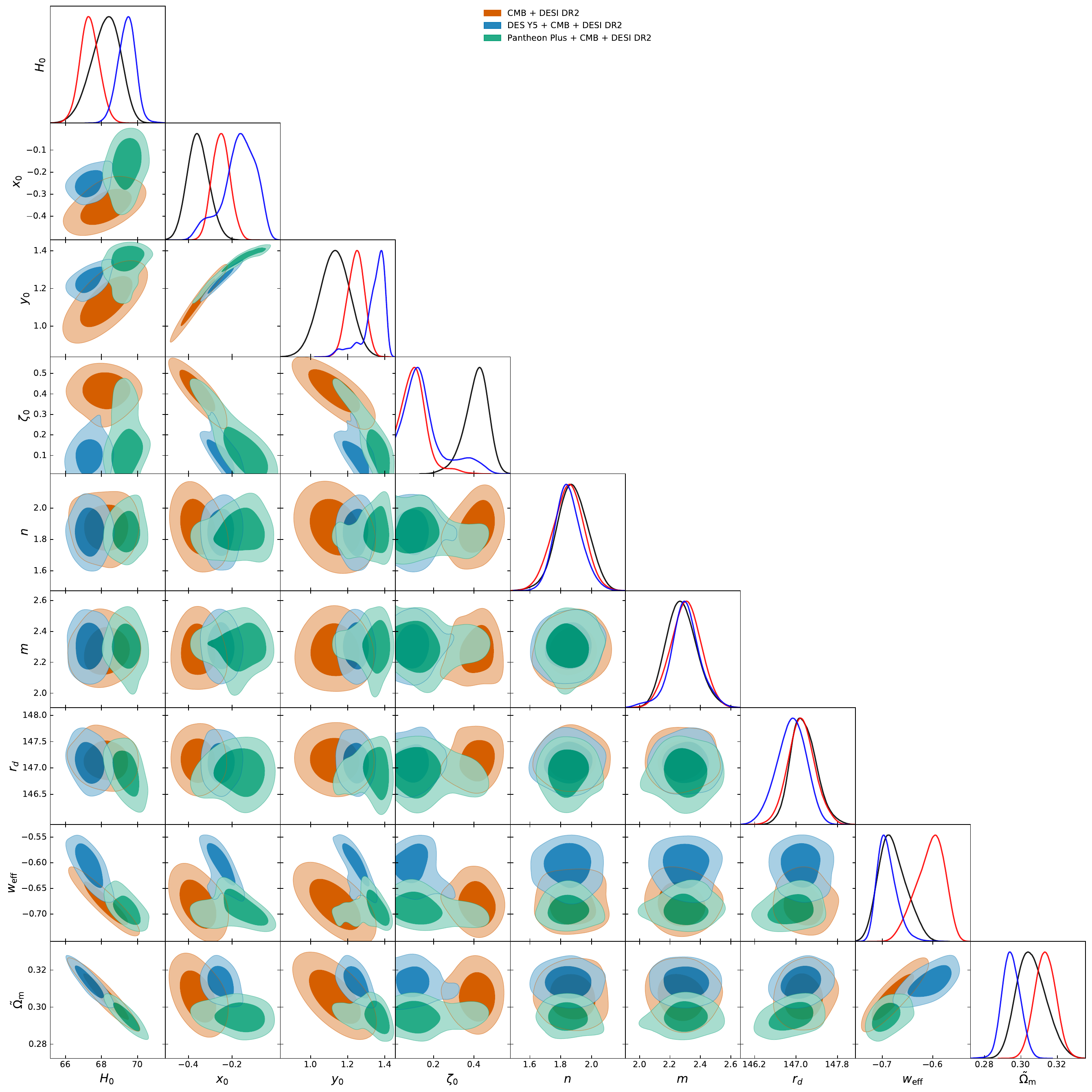}
	\caption{\textit{One- and two-dimensional marginalized posterior
	distributions for the parameters of Connection I obtained from
	Pantheon Plus+CMB+DESI DR2, DES Y5+CMB+DESI DR2, and CMB+DESI DR2.
	The filled contours correspond to the $68\%$ and $95\%$ credible
	regions.}}
	\label{fig:contoursI}
\end{figure*}

\begin{figure*}[t]
    \centering
    \includegraphics[width=\textwidth]{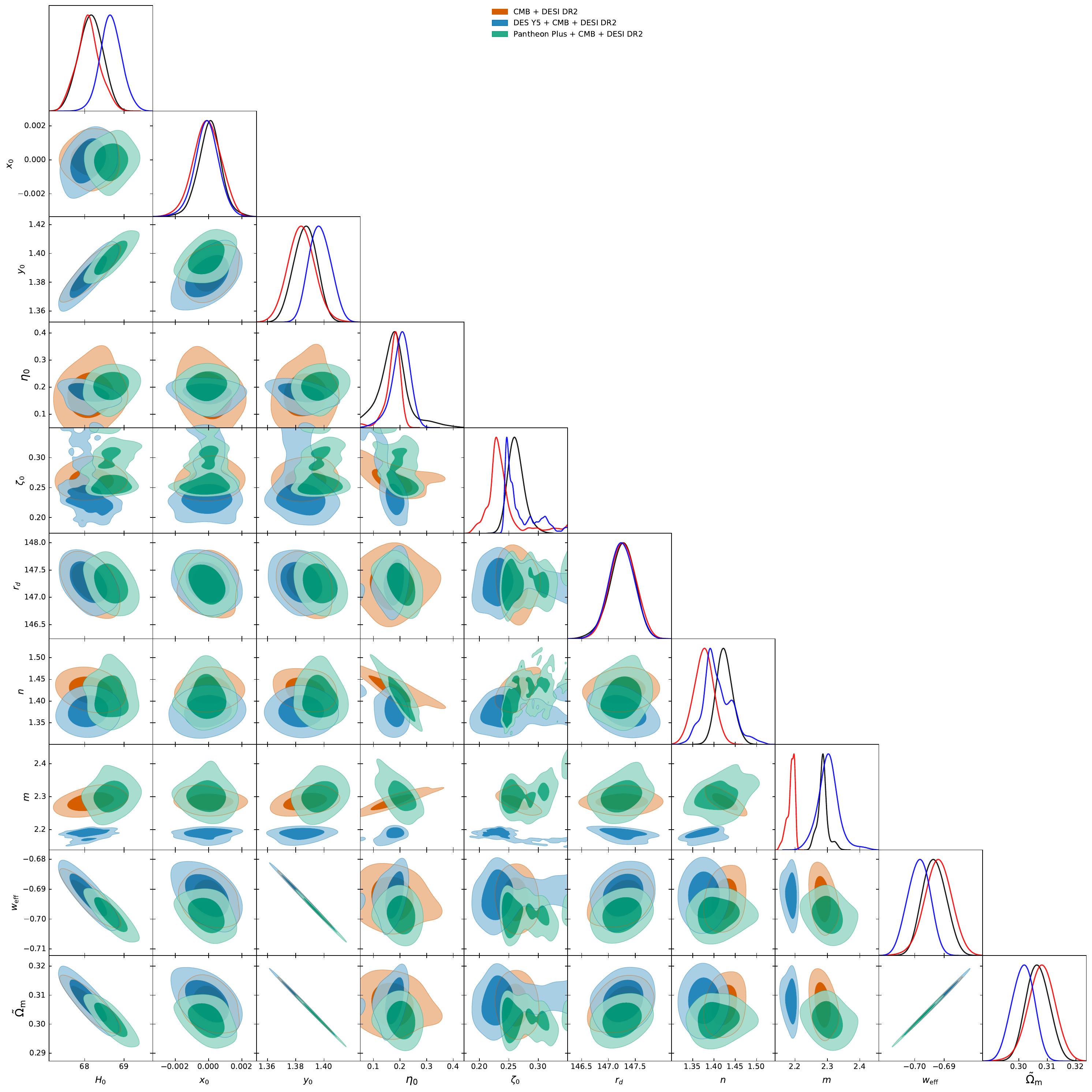}
    \caption{\textit{One- and two-dimensional marginalized posterior
	distributions for the parameters of Connection II obtained from
	Pantheon Plus+CMB+DESI DR2, DES Y5+CMB+DESI DR2, and CMB+DESI DR2.
	The filled contours correspond to the $68\%$ and $95\%$ credible
	regions.}}
    \label{fig:contoursII}
\end{figure*}

For Connection II, the three observational combinations lead to a much
more uniform set of constraints. We obtain
$H_0=68.67^{+0.26}_{-0.27}\ {\rm km\,s^{-1}\,Mpc^{-1}}$ from Pantheon
Plus+CMB+DESI DR2,
$H_0=68.09^{+0.29}_{-0.30}\ {\rm km\,s^{-1}\,Mpc^{-1}}$ from DES
Y5+CMB+DESI DR2, and
$H_0=68.15\pm0.30\ {\rm km\,s^{-1}\,Mpc^{-1}}$ from CMB+DESI DR2.
The normalized matter density remains close to
$\widetilde{\Omega}_{m0}\simeq0.30$ in all three cases, while
$r_d\simeq147.2$ Mpc. Moreover, the present scalar kinetic variable is
strongly constrained around $x_0\simeq0$, whereas
$y_0\simeq1.38-1.39$. The inferred effective equation of state is also
remarkably stable, with $w_{\rm eff,0}\simeq-0.69$ and uncertainties of order a few
$\times 10^{-3}$. Therefore, in contrast to Connection I, the main
late-time cosmological quantities in Connection II are only weakly
affected by whether Pantheon Plus, DES Y5, or no supernova sample is
included.

This behavior is directly reflected in the reconstructed histories shown in
Fig.~\ref{fig:evolutionII}. The three dataset combinations produce almost
overlapping central evolutions for
$\widetilde{\Omega}_{m}(z)$,
$\widetilde{\Omega}_{\rm DE}(z)$, and $w_{\rm eff}(z)$, with comparatively
narrow credible regions. The evolution again interpolates between an
approximately matter-dominated regime at higher redshift and the
late-time accelerated phase. Hence, although the affine connection enters
explicitly in the background equations of this branch, the observationally
allowed expansion history is tightly restricted and exhibits only a weak
dependence on the particular supernova compilation.

Finally, Connection III also leads to well-constrained late-time
cosmological parameters. For Pantheon Plus+CMB+DESI DR2 we find
$H_0=68.73\pm0.14\ {\rm km\,s^{-1}\,Mpc^{-1}}$ and
$\widetilde{\Omega}_{m0}=0.29^{+0.017}_{-0.016}$, while DES
Y5+CMB+DESI DR2 gives
$H_0=68.63^{+0.17}_{-0.15}\ {\rm km\,s^{-1}\,Mpc^{-1}}$ and
$\widetilde{\Omega}_{m0}=0.27^{+0.025}_{-0.014}$. For CMB+DESI DR2 we
obtain
$H_0=68.56^{+0.15}_{-0.14}\ {\rm km\,s^{-1}\,Mpc^{-1}}$ and
$\widetilde{\Omega}_{m0}=0.27^{+0.045}_{-0.030}$. In all three cases
$r_d$ remains close to $147.2$ Mpc. The present effective equation of
state lies in the interval $w_{\rm eff,0}\simeq-0.68$ to $-0.72$, while
the preferred values of the steep-potential parameters remain concentrated
around $m\simeq2.3$ and $n\simeq1.4$. Thus, the inferred value of $H_0$
is particularly stable in this branch, while the matter fraction shows a
somewhat stronger dependence on the dataset combination.

The reconstructed evolution for Connection III is displayed in
Fig.~\ref{fig:evolutionIII}. As in the previous two branches, the
observationally allowed histories exhibit the transition from a
matter-dominated regime towards late-time accelerated expansion. The
Pantheon Plus and DES Y5 reconstructions remain relatively close, whereas
the CMB+DESI DR2 combination allows somewhat greater freedom in the
relative matter and effective dark-energy fractions. Nevertheless, the
corresponding $w_{\rm eff}(z)$ histories remain mutually compatible within
their credible regions and approach values close to zero towards higher
redshift. Therefore, the additional geometrical contribution associated
with Connection III remains compatible with the combined low- and
high-redshift probes considered here, while producing an observational
behavior that is distinct from both Connections I and II.

Overall, all three symmetric-teleparallel connection branches provide
cosmological histories compatible with a matter-dominated epoch followed
by late-time accelerated expansion. However, the degree to which the
inferred background evolution depends on the observational dataset is
clearly connection dependent. Connection I exhibits the strongest
sensitivity to the choice of supernova compilation, most notably through
the shifts in $H_0$ and $w_{\rm eff,0}$. Connection II displays the most
stable reconstruction, with the three dataset combinations yielding
nearly coincident expansion histories, whereas Connection III presents an
intermediate behavior, with a very stable value of $H_0$ but somewhat
larger freedom in the reconstructed matter fraction. These differences
demonstrate that the choice of symmetric-teleparallel connection can leave
a measurable imprint on the inferred background cosmology, even when all
three branches remain compatible with the observational probes considered
here. A more detailed examination of the multidimensional parameter
constraints and their correlations is presented in the following
subsection.

\begin{figure*}[t]
    \centering
    \includegraphics[width=\textwidth]{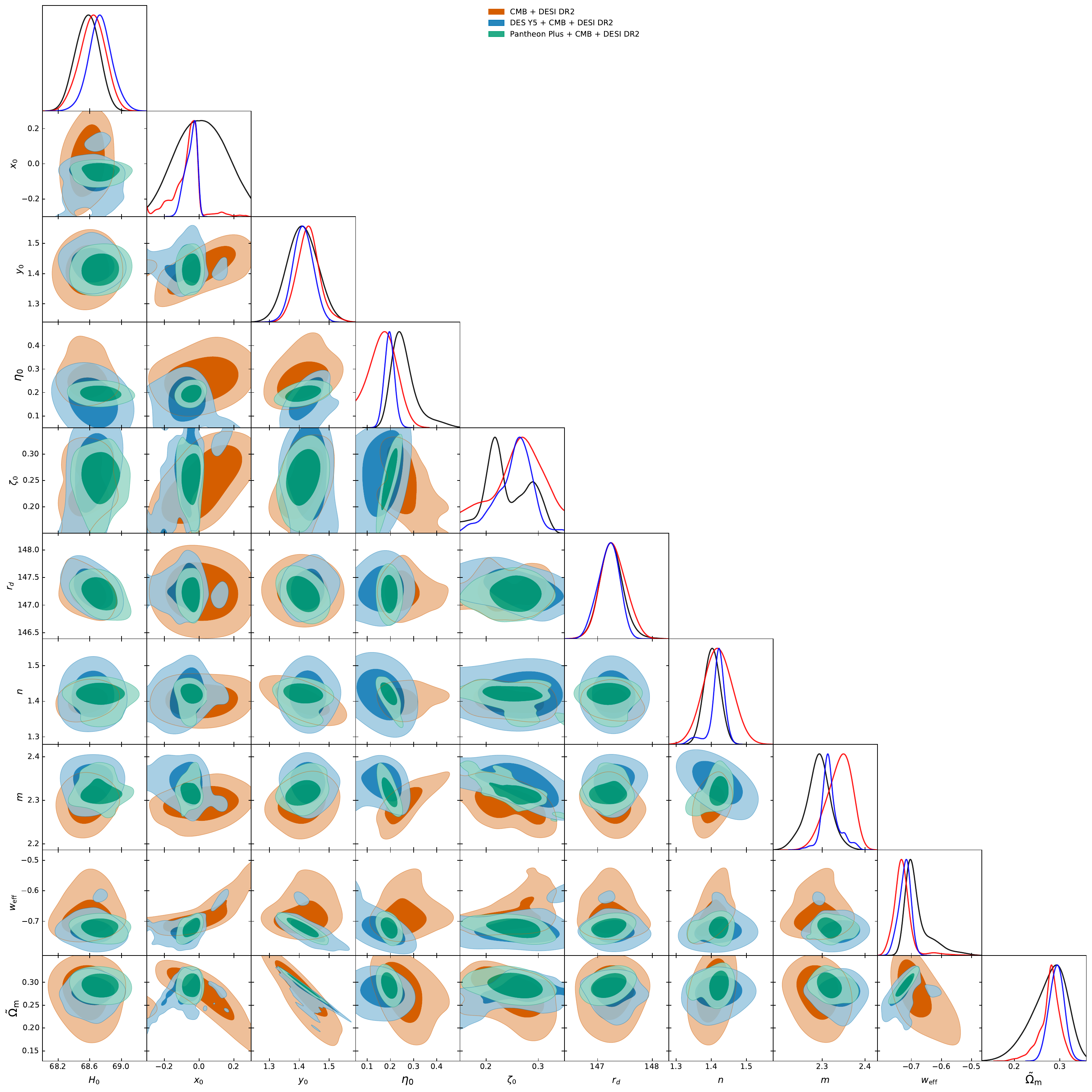}
    \caption{\textit{One- and two-dimensional marginalized posterior
	distributions for the parameters of Connection III obtained from
	Pantheon Plus+CMB+DESI DR2, DES Y5+CMB+DESI DR2, and CMB+DESI DR2.
	The filled contours correspond to the $68\%$ and $95\%$ credible
	regions.}}
    \label{fig:contoursIII}
\end{figure*}

\begin{table*}
	\centering
	\tiny
	\begin{tabular}{l c c c c c c}
		\hline
		\multicolumn{5}{c}{\bf Datasets: Pantheon Plus + CMB + DESI DR2}\\
		\hline
		\hline
		Model & $\chi_{ min}^{2}$ & $\Delta\chi_{min}^{2}=\chi_{min}^{2}-\chi_{ min,
\Lambda CDM}^{2}$ & $AIC$  &$\Delta{AIC}={AIC}-{ AIC}_{\Lambda CDM}$  & \\
		\hline
		{\bf Connection I:}
		& $1484.20$
		& $-1.18$
		& $1498.20$
		& $6.82$
         \\

		{\bf Connection II:}
		&$1488.81$
		& $3.43$
		& $1504.81$
		& $13.43$
        \\

		{\bf Connection III:}
		&$1490.00$
		& $4.62$
		& $1506.00$
		& $14.62$
        \\
		{\boldmath\bf $\Lambda$CDM:}
		&$1485.38$
		& $0$
		& $1491.38$
		& $0$
        \\

		\hline
        \multicolumn{5}{c}{\bf  Datasets: DES Y5 + CMB + DESI DR2}\\
		\hline
        \hline
		{\bf Connection I:}
		& $1645.77$
		& $-13.99$
		& $1659.77$
		& $-5.99$
         \\

		{\bf Connection II:}
		&$1659.63$
		& $-0.13$
		& $1675.63$
		& $9.87$
        \\

        {\bf Connection III:}
		&$1660.96$
		& $1.20$
		& $1676.96$
		& $11.20$
        \\
		{\boldmath\bf $\Lambda$CDM:}
		&$1659.76$
		& $0$
		& $1665.76$
		& $0$
        \\
        \hline
         \multicolumn{5}{c}{\bf Datasets: CMB + DESI DR2}\\
		\hline
        \hline
		{\bf Connection I:}
		& $9.31$
		& $-1.00$
		& $23.31$
		& $7.00$
         \\

		{\bf Connection II:}
		&$10.65$
		& $0.34$
		& $26.65$
		& $10.34$
        \\

        {\bf Connection III:}
		&$10.39$
		& $0.08$
		& $26.39$
		& $10.08$
        \\
		{\boldmath\bf $\Lambda$CDM:}
		&$10.31$
		& $0$
		& $16.31$
		& $0$
        \\
        \hline
		\hline
	\end{tabular}
	\caption{Statistical comparison of Connection I, Connection II, Connection III,
and the $\Lambda$CDM model using the $\chi^2_{min}$, $\Delta\chi^2_{min}$, AIC, and
$\Delta$AIC values obtained from MCMC analyses with the Pantheon Plus+CMB+DESI DR2,
DES Y5+CMB+DESI DR2, and CMB+DESI DR2 datasets.}
	\label{tab:modelcomparison}
\end{table*}

\subsection{Posterior distributions and model comparison}
\label{subsec:posterior_comparison}

We finally examine the multidimensional posterior distributions and assess
the statistical performance of the three connection branches relative to
$\Lambda$CDM. The one- and two-dimensional marginalized posterior
distributions are presented in
Figs.~\ref{fig:contoursI}-\ref{fig:contoursIII}. They complement the
one-dimensional constraints discussed above by displaying the correlations
and degeneracies among the cosmological parameters, the scalar-field
variables, and, for Connections II and III, the variables associated with
the affine connection.

For Connection I, shown in Fig.~\ref{fig:contoursI}, the posterior
distributions exhibit a noticeable dependence on the adopted supernova
sample. This is particularly evident in quantities controlling the
late-time expansion, consistently with the shifts in $H_0$,
$\widetilde{\Omega}_{m0}$, and $w_{\rm eff,0}$ found in
Sec.~\ref{subsec:constraints_evolution}. The Pantheon Plus and
CMB+DESI DR2 constraints show substantial overlap in several parameter
directions, whereas the inclusion of DES Y5 shifts the preferred region
and modifies a number of parameter degeneracies. This behavior confirms
that Connection I possesses comparatively large phenomenological freedom
at the background level and can accommodate different preferred
late-time expansion histories depending on the supernova compilation
employed.

The posterior distributions for Connection II are displayed in
Fig.~\ref{fig:contoursII}. In this case the three dataset combinations
select considerably more similar regions of parameter space. In
particular, the late-time cosmological quantities show only moderate
shifts when the supernova compilation is changed, in agreement with the
nearly overlapping reconstructed histories of Fig.~\ref{fig:evolutionII}.
The posterior distributions also confirm the strong localization of
$x_0$ around zero, while the remaining scalar and connection variables
are confined to comparatively narrow regions. Hence, Connection II
exhibits a substantially more dataset-independent background
phenomenology than Connection I.

Connection III, whose posterior distributions are presented in
Fig.~\ref{fig:contoursIII}, displays an intermediate behavior. The three
observational combinations remain broadly compatible, although the
CMB+DESI DR2 case, which does not contain supernova information, allows
larger regions for some of the background variables. At the same time,
the steep-potential parameters remain concentrated around
$m\simeq2.3$ and $n\simeq1.4$, while the preferred values of $H_0$ are
remarkably stable among the three dataset combinations. Thus,
Connections II and III both exhibit a comparatively robust observational
behavior, although their multidimensional posterior structures are
different, reflecting the different manner in which the affine connection
enters their respective background equations.

In order to quantify the relative statistical performance of the three
connection branches, we compare their minimum chi-square values and the
Akaike information criterion (AIC), defined as
\begin{equation}
{\rm AIC}
=
\chi^2_{\min}+2k,
\label{AIC}
\end{equation}
where $k$ is the number of free parameters of the corresponding model.
We further define
\begin{equation}
\Delta\chi^2_{\min}
=
\chi^2_{\min}-\chi^2_{\min,\Lambda{\rm CDM}},
\label{Deltachi2a}
\end{equation}
and
\begin{equation}
\Delta{\rm AIC}
=
{\rm AIC}-{\rm AIC}_{\Lambda{\rm CDM}},
\label{DeltaAIC}
\end{equation}
such that negative values favor the scalar-nonmetricity model relative to
$\Lambda$CDM, while positive values favor $\Lambda$CDM. The corresponding
results are summarized in Table~\ref{tab:modelcomparison}.

For Pantheon Plus+CMB+DESI DR2, Connection I achieves a slightly smaller
minimum chi-square than $\Lambda$CDM, with
$\Delta\chi^2_{\min}=-1.18$. This improvement in the fit is nevertheless
too small to compensate for the additional model parameters, resulting in
$\Delta{\rm AIC}=6.82$. Connections II and III give
$\Delta\chi^2_{\min}=3.43$ and $4.62$, respectively, together with
$\Delta{\rm AIC}=13.43$ and $14.62$. Hence, for this dataset combination
the additional freedom of the scalar-nonmetricity scenarios is not
favored by the AIC, and $\Lambda$CDM provides the more economical
description of the data.

A markedly different behavior is obtained for the
DES Y5+CMB+DESI DR2 combination. In this case Connection I reduces the
minimum chi-square by
$\Delta\chi^2_{\min}=-13.99$ relative to $\Lambda$CDM. The improvement
is sufficiently large to overcome the AIC penalty associated with the
additional parameters, yielding
$\Delta{\rm AIC}=-5.99$. Thus, among all the scalar-nonmetricity cases
and dataset combinations considered here, this is the only one for which
the AIC favors the modified-gravity scenario over $\Lambda$CDM.
Connection II produces only a very small improvement in the minimum
chi-square,
$\Delta\chi^2_{\min}=-0.13$, which is insufficient to compensate for its
additional parameters and gives $\Delta{\rm AIC}=9.87$. Connection III
instead yields $\Delta\chi^2_{\min}=1.20$ and
$\Delta{\rm AIC}=11.20$.

For CMB+DESI DR2, the four models achieve very similar minimum chi-square
values. In particular, Connections I, II, and III give
$\Delta\chi^2_{\min}=-1.00$, $0.34$, and $0.08$, respectively.
Consequently, there is little difference in their maximum-likelihood
description of these data. Once the additional free parameters are taken into
account, however, the corresponding values become
$\Delta{\rm AIC}=7.00$, $10.34$, and $10.08$, respectively. Hence,
$\Lambda$CDM is again favored according to the AIC   for this
dataset combination.

The model comparison therefore reveals an important connection dependence
that goes beyond the differences already seen in the reconstructed
background histories. All three branches can accommodate the observed
late-time expansion, but they do so with different degrees of parameter
freedom and different responses to the observational datasets. Connection I
is the most sensitive to the adopted supernova compilation and exhibits
the largest phenomenological flexibility. This flexibility does not improve
its statistical performance for Pantheon Plus or for CMB+DESI DR2 alone,
once the additional parameters are penalized. Remarkably, however, for the
DES Y5 combination it leads to a reduction of almost fourteen units in
$\chi^2_{\min}$ and to $\Delta{\rm AIC}<0$, such that the improvement in
fit outweighs the complexity penalty. Connections II and III, in contrast,
display considerably more stable posterior constraints and reconstructed
histories across the three dataset combinations, but this robustness is
accompanied by no comparable improvement in the best-fit likelihood.

Hence, the observational impact of the symmetric-teleparallel connection
is twofold. It affects both the reconstructed expansion history and the
statistical response of the model to different combinations of cosmological
data. In particular, the fact that the same scalar-nonmetricity action and
the same steep potential lead to distinct posterior structures and model
comparison results for the three allowed connections demonstrates that the
choice of affine connection is not observationally innocuous at the
background level. This constitutes one of the main results of the present
analysis.

\section{Conclusions}
\label{Conclusion}

The geometric formulation of gravity based on nonmetricity has emerged as
an important alternative framework for exploring gravitational physics
beyond General Relativity. In particular, $f(Q)$ gravity has demonstrated
that symmetric teleparallel geometry can lead to a rich cosmological
phenomenology while preserving a conceptually distinct description of the
gravitational interaction. Scalar-nonmetricity theories provide a further
natural extension, in close analogy with scalar-tensor and scalar-torsion
gravity, by allowing a scalar degree of freedom to couple nonminimally to
the nonmetricity scalar. An especially interesting feature of this
framework is that, unlike in the usual metric formulation, cosmological
symmetry does not uniquely determine the affine connection. Even for a
spatially flat Friedmann-Robertson-Walker background, three inequivalent
symmetric-teleparallel connection branches are allowed. Consequently, the
choice of connection may acquire physical relevance and, in principle,
leave observable imprints on the cosmological evolution.

In the present work we investigated this possibility by performing a
systematic observational analysis of scalar-nonmetricity cosmology for all
three connection branches. We considered a steep scalar-field potential,
$U(\phi)=U_0\exp(\alpha\phi^n)$, together with a power-law nonminimal
coupling, $f(\phi)=\beta\phi^m$. Such a choice is particularly interesting
because the steepness of the potential, the nonminimal coupling, and the
connection contribution can compete in determining the late-time evolution.
After deriving the corresponding background equations in a form suitable
for numerical integration, we confronted the three branches with
combinations of Pantheon Plus or DES Year 5 Type Ia supernovae, DESI DR2
BAO measurements, and compressed CMB information. This allowed us to
constrain the model parameters, reconstruct the cosmological histories, and
compare the statistical performance of the three connections with that of
$\Lambda$CDM. In this way, our analysis directly addresses whether the
freedom associated with the symmetric-teleparallel connection can be
probed observationally.

Our results show first of all that all three connection branches can
provide viable background cosmological histories. In every case the
observationally allowed evolution exhibits the required transition from a
matter-dominated regime at higher redshift to an effective dark-energy-dominated
phase at late times, accompanied by the onset of accelerated
expansion. Nevertheless, the three branches do not respond to the data in
the same way. Connection I displays the strongest dependence on the
supernova compilation. In particular, the inferred values of $H_0$ and
$w_{\rm eff,0}$ change noticeably when Pantheon Plus is replaced by DES
Y5, and this behavior is clearly reflected in the reconstructed histories
and posterior distributions. Hence, although Connection I is the simplest
branch at the level of the background connection structure, it possesses
considerable phenomenological freedom in the late Universe.

Connection II exhibits a markedly different behavior. The three dataset
combinations lead to very similar values of the main cosmological
quantities, with $\widetilde{\Omega}_{m0}\simeq0.30$,
$r_d\simeq147.2$ Mpc, and $w_{\rm eff,0}\simeq-0.69$. The reconstructed
histories are almost overlapping and the corresponding posterior regions
are comparatively narrow. Thus, despite the explicit appearance of the
affine connection in the background equations, the observationally allowed
cosmology of this branch is highly restricted and remarkably insensitive
to the particular supernova compilation. Connection III presents an
intermediate situation. Its inferred value of $H_0$ is particularly stable
among the three observational combinations, while somewhat larger freedom
is allowed in the reconstructed matter and effective dark-energy
fractions. At the same time, the steep-potential parameters are consistently
localized around $m\simeq2.3$ and $n\simeq1.4$. These differences between
the three branches already demonstrate that the connection choice affects
the way in which a given scalar-nonmetricity action is mapped onto the
observationally allowed cosmic expansion.

The model comparison strengthens this conclusion. For
Pantheon Plus+CMB+DESI DR2, Connection I slightly improves the minimum
chi-square relative to $\Lambda$CDM, but the improvement is not sufficient
to compensate for its additional parameters, while Connections II and III
are also disfavored according to the AIC. A qualitatively
different result arises for DES Y5+CMB+DESI DR2. In this case Connection I
achieves
$\Delta\chi^2_{\min}=-13.99$ and
$\Delta{\rm AIC}=-5.99$, and is therefore favored over $\Lambda$CDM
according to the AIC among the models considered. Neither Connection II nor
Connection III exhibits a comparable improvement. Finally, for CMB+DESI
DR2 alone the minimum chi-square values of all four cosmologies are very
similar, and the AIC consequently favors the more economical
$\Lambda$CDM description. Hence, the additional freedom of
scalar-nonmetricity gravity is not generically rewarded by the data.
However, the fact that one connection branch can exploit this freedom to
improve substantially the fit to a specific modern supernova dataset,
whereas the other two cannot, is particularly noteworthy. It indicates
that the choice of symmetric-teleparallel connection affects not only the
form of the cosmological equations, but also the statistical response of
the theory to observations.

There are several directions in which the present analysis can be extended.
A natural next step is to move beyond the homogeneous background and study
linear cosmological perturbations for the three connection branches,
including the evolution of matter perturbations, the effective
gravitational coupling, gravitational slip, and structure-growth
observables. This would allow the models to be tested against redshift-space
distortions, weak lensing, large-scale structure, and $S_8$ measurements,
and could reveal additional differences between connections that remain
partly degenerate at the background level. It would also be important to
incorporate the full CMB likelihood, investigate broader classes of
scalar-field potentials and nonminimal couplings, and complement the AIC
analysis with Bayesian model comparison. Future standard-siren,
large-scale-structure, and next-generation supernova and BAO measurements
will further improve the ability to distinguish between the different
branches. More generally, identifying whether observations can favor a
particular symmetric-teleparallel connection may provide valuable guidance
for understanding which geometrical realization of nonmetricity gravity is
physically relevant. The results obtained here provide a first step in this
direction and show that the connection freedom inherent in
scalar-nonmetricity gravity can translate into genuinely distinguishable
cosmological phenomenology.

\end{document}